\documentclass{sig-alternate-05-2015}
\usepackage{epsfig,endnotes}
\usepackage[inline]{enumitem}
\usepackage{microtype}
\usepackage{balance}
\graphicspath{ {./img/} {./images/} }
\DeclareGraphicsExtensions{.pdf,.jpeg,.jpg,.png}
\usepackage{url}

\usepackage{enumerate}

\usepackage{graphicx}

\begin{document}

%

\title{\Large \bf Towards Decentralised Resilient Community Cloud Infrastructures}

\author{
{Arjuna Sathiaseelan, Mennan Selimi, Carlos Molina, Adisorn Lertsinsrubtavee}\\
University of Cambridge
\and
{Leandro Navarro, Felix Freitag}\\
Universitat Polit\`ecnica de Catalunya (UPC)
\and
{Fernando Ramos}\\
University of Lisbon
\and
{Roger Baig}\\
Guifi.net Foundation
} 

\maketitle



%


\maketitle
\begin{abstract}
Recent years have seen a trend towards decentralisation - from initiatives on decentralized web to decentralized network infrastructures (e.g community networks). In this position paper, we present an architectural vision for decentralising cloud service infrastructures. Our vision is on the notion of community cloud infrastructures on top of decentralised access infrastructures i.e. community networks, using resources pooled from the community. Our architectural vision takes into consideration some of the fundamental challenges of integrating the current state of the art virtualisation technologies such as Software Defined Networking (SDN) into community infrastructures which are highly unreliable. Our proposed design goal is to include lightweight virtualization and fault tolerance mechanisms into the architecture to ensure sufficient level of reliability to support critical applications. 
\end{abstract}




\section{Introduction}
\begin{figure}
    \centering
    \includegraphics[width=\linewidth]{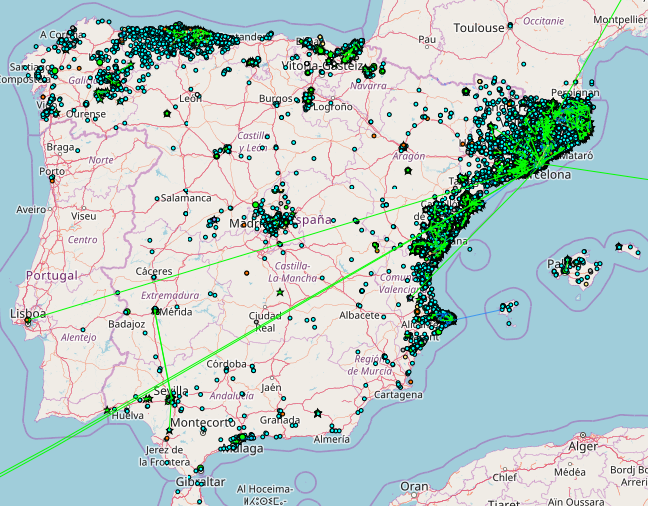}
    \caption[Guifi.net nodes]{Guifi.net nodes and links}
    \label{fig:guifi_nodes}
\end{figure}

The Internet, which started off as a decentralised network, has over the years become increasingly centralised with a staggering percentage of communications flowing through access and service infrastructures controlled by a set of corporations who are (naturally) economically motivated. Such centralisation of infrastructures mean, for the other three billion (who are yet to go online) to get connected and realise the benefits of the Internet, they are dependent on existing big players. Decentralisation is hence a form of empowerment, allowing fairer participation, where smaller players (e.g. local stakeholders) can make things happen on their own. To expedite the process of connecting the remaining three billion, we need to enable and support decentralised bottom up initiatives (such as community owned infrastructures and services) that are complimentary to existing centralised services. 
Recent years have witnessed this growing trend of decentralisation of infrastructures through the rise of the Community Networks (CNs). The reluctance of network operators to provide wired and cellular infrastructures to rural/remote areas as well as the need for affordable connectivity in urban areas have led to several bottom up community led initiatives to build large-scale, self-organized, decentralized CNs. In such CNs, communities pool their resources to build and govern their own access and service infrastructures. There are several examples of successful CNs, most notably Guifi.net~\cite{Baig2015150}. Guifi.net which started off in Barcelona in 2004 has currently a total of $33,800$ nodes, accounting for $52,354$ WiFi links making a total length of $58,612$ km. Figure \ref{fig:guifi_nodes} shows as example the nodes and links of Guifi.net in the city of Barcelona.

With many examples around the world, CNs grow organically without any strict planning and hence face several technology challenges. They are heterogeneous in nature with a high level of node and network diversity (including different topologies), raising problems associated with resource management, instability and unavailability. 

\begin{figure}
    \centering
    \includegraphics[width=0.9\linewidth] {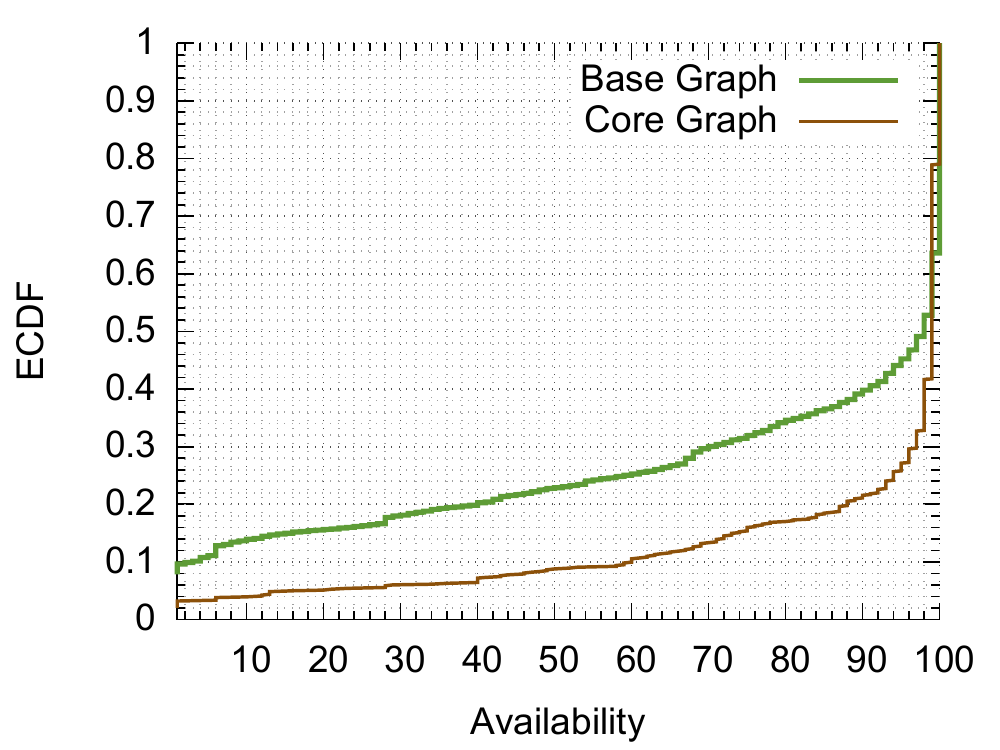}
    \caption[Node availability]{Node availability}
    \label{fig:guifi_ava}
\end{figure}

\textbf{Infrastructure:} The infrastructure in the CN such as Guifi.net is highly unreliable and heterogeneous \cite{VegaCN}. Devices and the network are very heterogeneous compared to the Data Centers (DCs) where they are very homogeneous. The strong heterogeneity is due to the diverse capacity of nodes and links, as well as the asymmetric quality of wireless links. Employed technologies in the CNs vary significantly, ranging from very low-cost, off-the-shelf wireless (WiFi) routers, home gateways, laptops to different type of antennas (e.g., parabolic) and optical fibre equipment. These devices forming the \emph{community clouds} are co-located in either users homes or within other infrastructures distributed in the CNs. The unreliability of current CN devices and the network limits their usability in critical applications that demand predictable performance or can assure Quality of Service (QoS). 
For instance we can examine the availability of the Guifi.net nodes in Barcelona as shown in Figure \ref{fig:guifi_ava}. We have conducted experiments to measure reachability, represented in Figure \ref{fig:guifi_ava} in a ECDF plot of the availability of the nodes. We assume that a node is available if it is reachable. We can see that $40\%$ of the \emph{base-graph} nodes (i.e., client nodes) are reachable from the network $90\%$ or less of the time. The situation seems to be better with the \emph{core-graph} nodes, which are the most stable part of the network (i.e., $20\%$ of the \emph{core-graph} nodes have an availability of $90\%$ or less). Core-graph nodes have higher availability because they comprise the backbone of the network and connect different administrative zones. The successful operation of the base-graph nodes depends on the core-graph nodes.


\textbf{Topology:} The network topology in a wireless CN such as Guifi.net is organic and different with respect to conventional ISP (Internet Service Provider) networks. Guifi.net is composed of numerous distributed CNs and they represent different types of network topologies. The overall topology is constantly changing and there is no fixed
topology as in the DC environment. The Guifi.net network shows some typical patterns from the urban networks (i.e., meshed networks) combined with an unusual deployment, that do not completely fit neither with organically grown networks nor with planned networks \cite{Vega2012}. This implies that a solution (i.e., algorithm) that works in a certain topology might not work in another one.

\begin{figure}
    \centering
    \includegraphics[width=0.9\linewidth] {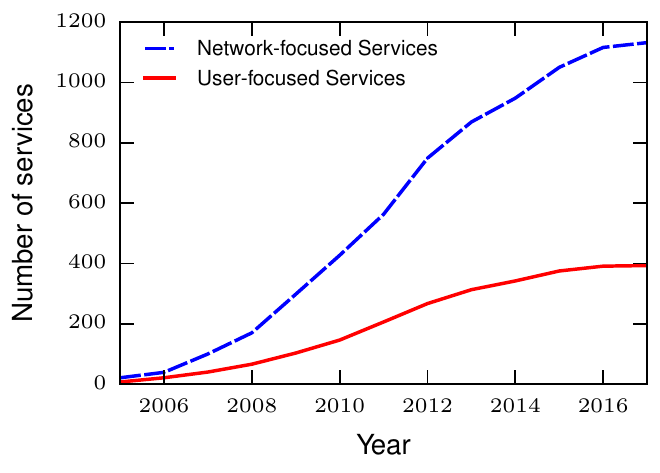}
    \caption[Number of user and network services]{Number of user and network services}
    \label{fig:services}
\end{figure}

\textbf{Services:} In the Guifi.net CN, Internet cloud services have equivalent local alternatives that are owned and operated at the community level. There are two type of services in the network: network-focused and user-focused services. Figure \ref{fig:services} depicts the evolution in the number of instances of user and network-focused services during the last 10 years. Considering that network management is of interest to all users in the network (i.e., to keep the network up and running), Figure \ref{fig:services} reveals that service instances related to the network operation outnumber the local services intended for end-users. Moreover, one of the most frequent of all the services, whether user-focused or network-focused, are the proxy services \cite{PAM17}. Proxies act as free shared gateways to the Internet for the CN users. Specifically for the user-focused services, the percentage of Internet access services (i.e., proxies and tunnel-based) is higher than $55\%$, confirming that the users of Guifi.net are typically interested in accessing the Internet \cite{Selimi2015}. Further, other important services are web hosting, data storage, VoIP, and video streaming.

Operating in this large and constantly changing environment requires the deployment of resilient service infrastructures that can efficiently react to diverse changes. Enabling resilient service infrastructures are crucial to the economic sustainability of CNs as it creates new economic opportunities for the community e.g. incentivising local service providers to utilise, contribute and maintain the CN infrastructure to provide new services. 

These results suggest that, with its current reliability Guifi.net can hardly be used to support critical applications in terms of strict performance or availability guarantees. 
This is the case for critical local sensing and control services, 
such as smart power grids that take use two-way communication to integrate supply to end-device consumption to implement dynamic energy prices and on-demand generation. Other examples are IoT applications used in smart cities like for road traffic management, such as street-line parking that relies on two way communication between sensors and car-drivers.



Though both publicly available commercial clouds (e.g. Amazon AWS or Windows Azure) and community clouds (e.g. inside Guifi.net) are based on the fundamental idea of resource sharing through virtualisation techniques, they differ in several aspects. For instance, public cloud services are owned and managed by a central authority and have highly reliable infrastructure. Conversely, community clouds lack central ownership and management and are deployed at the edge of the Internet, normally close to the users. These particularities of local community clouds make them attractive to several applications for example, they are free from vendor lock-in, lower exposure of personal data to third parties, no reliance on good Internet connectivity, and lower latency.

The task of building resilient local community clouds on the basis of unreliable CNs is challenging. In this position paper, we layout our architectural vision where we argue that the problem of resilience can be solved with the assistance of Software Defined Network (SDN), Network Function Virtualisation (NFV) and lightweight OS virtualisation technologies in combination with monitoring techniques. The idea is to use such technologies to build fault tolerance mechanisms based on replication, redundancy, relocation,  decentralisation and dynamic reconfiguration performed on the basis of monitored metrics. In the subsequent sections, we explain the general ideas and discuss some preliminary work that we have conducted in this direction.

\section{Community Clouds}
\label{clouds}

CN clouds (i.e., micro-clouds) are built on top of the CNs. In this model, a cloud is deployed closer to CN users and other existing infrastructure. The community cloud model is different from Fog computing, which extends cloud computing by introducing an intermediate layer between devices and data centers. Community clouds take the opposite track, by putting services closer to consumers, so that no further or minimal action takes place in the Internet. In that way, fragile Internet connectivity is no longer a key issue, and personal data and therefore privacy can be better preserved. They are deployed over a single or set of user nodes, and comparing to the public clouds they have a smaller scale, so one still gets high performance due to locality and control over service placement \cite{Selimi2017}. The infrastructure of Community Clouds consists
of mesh routers, mesh clients and optionally gateways (i.e., proxies) typically deployed on every home.

The routers (i.e., outdoor routers) communicate with each other via radio transmissions and employ special-purpose routing protocols as BMX6\cite{Neumann2015308}. Mesh clients access the network via one of the routers, while gateways provide connectivity to the Internet. Client nodes consists of home gateways, laptops or desktop PCs.

\textbf{Cloudy:} In order to foster the adoption and transition of the community cloud environment, we have developed Cloudy~\cite{Cloudy15}, a community cloud software distribution, which contains the infrastructure, platform and application services of the community cloud system. Cloudy is the core software of our community clouds, because it unifies the different tools and services of the cloud system in a Debian-based Linux distribution. Cloudy is open-source and can be downloaded from public repositories\footnote{https://github.com/Clommunity/}. Cloudy's main components can be considered a layered stack, with services residing both inside the kernel and at the user level. Services are running on Docker and LXC containers.

\section{Architecture}
\label{architecture}

\begin{figure}[t]
  \centerline{\includegraphics[angle=0,origin=c,scale=0.50]{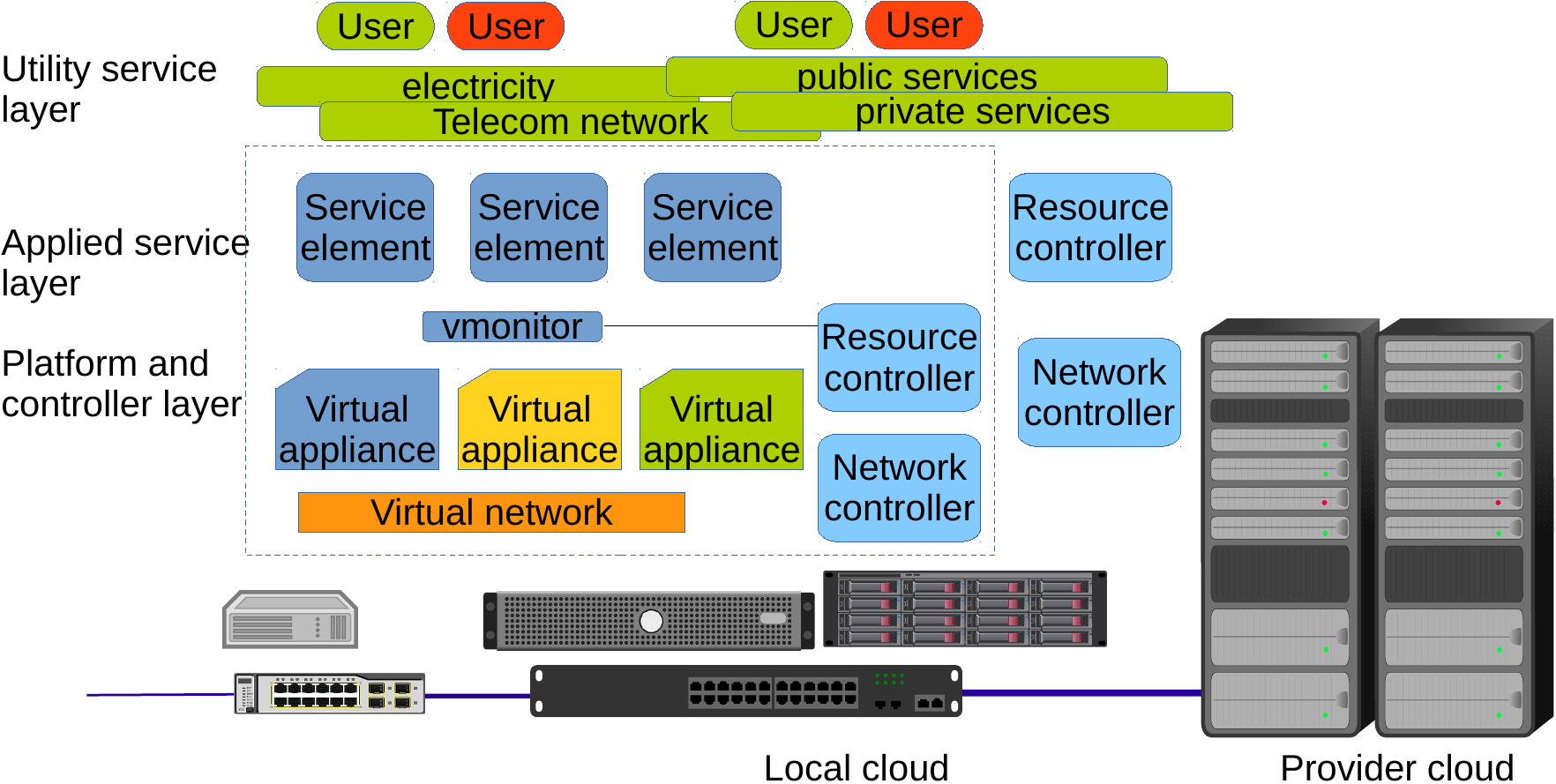}}
  \caption[Architecture of a slice]{Architecture of a slice}
  \label{fig:architecture_slice}
\end{figure}

A resilient community cloud infrastructure should allow the contribution of networking, computing, and storage resources, and the allocation of a fraction of these resources (a slice) to services composed by a set of application service-elements consuming these resources. A \textbf{resource slice} (see Figure~\ref{fig:architecture_slice})  is an integrated collection of resources with its own virtualised network, network controller, network functions, integrated with its set of application service-elements. The resource slice is exposed to external users' load and a set of varying environmental issues such as asymmetric or unidirectional paths, network partitions, component failures and congestion; that require a reactive and adaptive behaviour.

To enable efficient resource pooling of infrastructures both devices and networks, we will exploit the recent advances in lightweight OS virtualisation, NFV and SDN to build a resilient decentralised cloud infrastructure.

Advances in NFV allow services, computation or functions to be orchestrated or placed and executed in any part of a network, thus removing the need for (costly) custom built hardware boxes. The advances in NFV coupled with the advances in lightweight OS virtualisation technologies such as Docker or Unikernels~\cite{madhavapeddy2013unikernels}, services can be made entirely mobile and be potentially migrated between two nodes in a network at a very low cost. Considering that communities pool their (rather limited) resources, enabling flexible orchestration of services~\cite{picasso17} i.e by allowing services to be instantiated when and where needed as well as supporting (live) migration between nodes at a lower cost, is indeed a boon for decentralised community clouds. 

Once services have been initiated, SDN~\cite{Kreutz15} can be used for: 
\begin{enumerate*}[label=\alph*)]
\item Service chaining network functions by configuring routes in switches based on policies.
\item Modifying packet headers appropriately to support such service chaining.
\item Implementing load balancing algorithms avoiding the need for service providers/operators to manually install traffic splitting rules or use custom load balancing solutions.
\end{enumerate*}

CNs introduce specific challenges in terms of trust, reliability and cooperation inducing issues such as asymmetric paths, unidirectional paths, network partitions etc. that make implementing SDN extremely challenging. The challenge we face is to build a resilient communication infrastructure out of unreliable components. The problem can be addressed either with support for asynchronous control in SDN~\cite{Dimos13} (preload of rules in proxy controllers), or with the help of fault forecasting techniques complemented with fault tolerant techniques. That allows to handle gracefully situations that are difficult or too costly to anticipate and prevent~\cite{Algirdas2004} in an ideal environment.

The goal of our architecture is to design a reliable service on the basis of these unreliable components taking advantage of their inherent high decentralisation, cutting edge technology provided by SDNs, fault tolerant techniques and sophisticated cloud management algorithms. 
To address the problem we will take advantage of the logical centralization management offered by SDNs and of their flexibility in the virtualisation and dynamic relocation of network functions. We will also take advantage of the high level of connectivity (mesh topology) normally exhibited by community networks. The challenge is to develop techniques and algorithms that allows cloud service providers use SDN controllers to reconfigure their communication network dynamically in response to detected outages, ideally, through anticipation or pre-loading. 

\begin{figure}[t]
 \centerline{\includegraphics[angle=0,origin=c,scale=0.28]{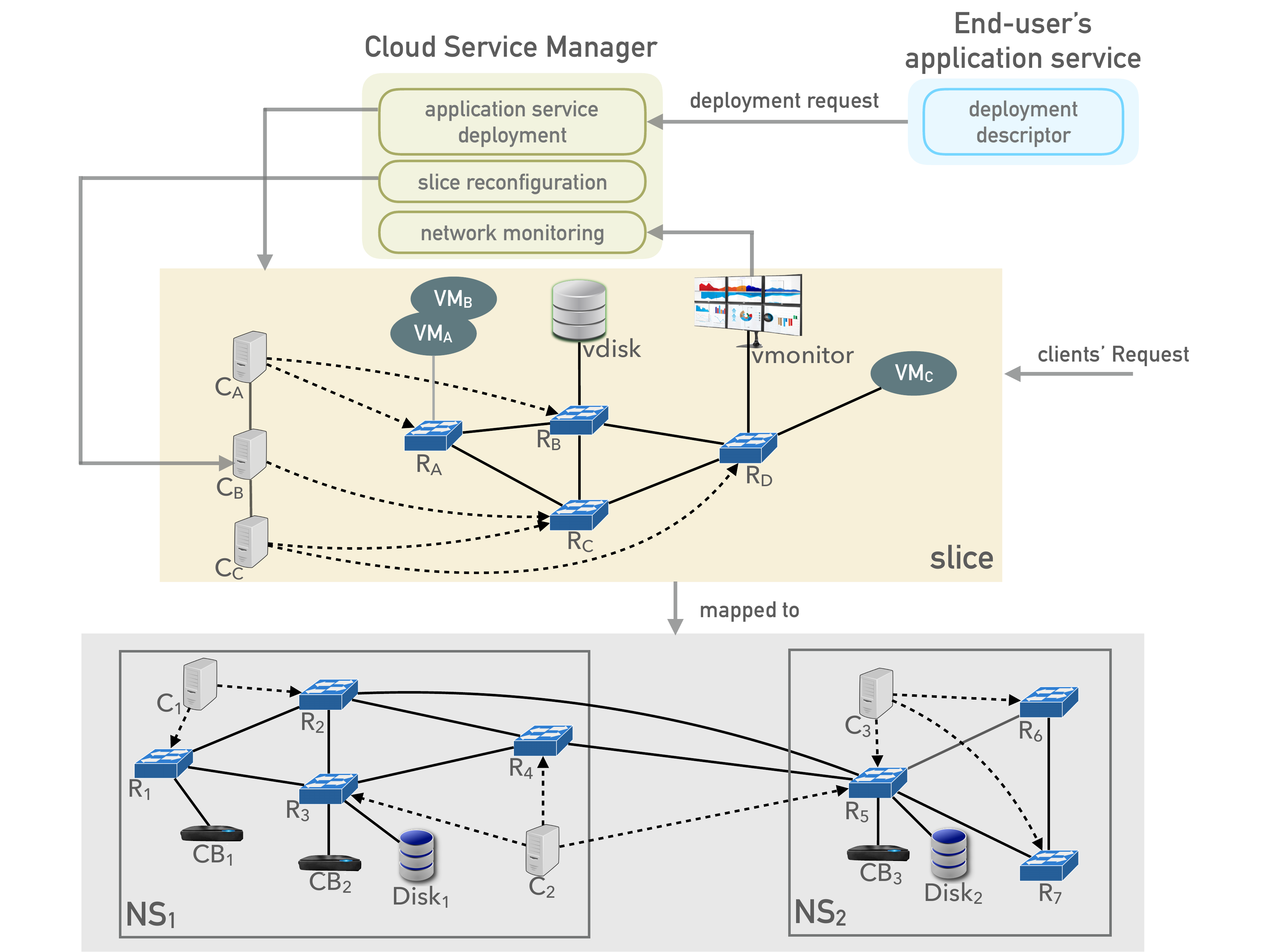}}
 \caption{Architecture of a resilient local community cloud}
\label{fig:appServiceDeployment}
\end{figure}

Our architecture integrates the resources offered by community networks with SDN components in a manner that third parties can use resource slices and offer them as cloud services to other parties to deploy their application services, under agreed upon QoS guarantees. Our proposed architecture can provide safety guarantees that the functionality of the applications service running on top is never compromised despite failures of individual components (for example, links) of the underlying infrastructure. Likewise, it can provide liveness properties that ensure that the underlying infrastructure will eventually recover and continue to deliver its service (for example, after a bounded reconfiguration time) despite a bounded number of network related failures.

\subsection{Architectural components}  
Figure~\ref{fig:appServiceDeployment} depicts a preliminary view of the architecture we propose. Note that this is only a conceptual view to illustrate our ideas. Figure~\ref{fig:appServiceDeployment} shows two independent substrate networks $NS_1$ and $NS_2$ (in principle, there might be more). We assume that they are unreliable community networks. $R_1$, $R_2$, etc., are switches, whereas $C_1$, $C_2$, etc., are controllers used for manipulating the switches. $CB_1$ and $CB_1$ are community boxes (running Cloudy) with computation resources capable of hosting virtual machines. Though not shown in the figure, community boxes include communication, computation and storage resources and are normally used by users to gain access to the community network. Note that in pursuit of reliability, some controllers might be replicated, like $C_2$ and $C_3$ in relation to $R_5$. In addition to communication, the substrate networks independently offer computation ($CB_1$ and
$CB_2$) and storage ($disk_1$ and $disk_2$) to Cloud Service Managers (CSMs) (only one of them in shown) by means
of resource virtualisation. The CSM is in full control of the virtualised resources and offers local cloud services to end users with QoS guarantees in spite of the unreliability exhibited by the substrate networks. The CSM is responsible for monitoring the pooled computational and storage resources and reacting to eventualities by means of reconfiguration. The CSM has a set of functions that enable the orchestration of resources. In Figure~\ref{fig:appServiceDeployment}, these functions are aggregated as \emph{application service deployment}, \emph{slice reconfiguration} and \emph{network monitoring} functions. These functions help the Cloud Service Manager to deal with additional complexity that fault-tolerance mechanisms inevitably introduce.

As depicted in Fig.~\ref{fig:appServiceDeployment}, an end user requests (\emph{deployment request}) the deployment of an application
service (\emph{End-user's application service}) - the resources and the QoS needed are specified
in a deployment descriptor
The request
might be accepted or rejected by the CSM. If the request is accepted, the CSM
via the \emph{application service deployment} function instantiates a resource slice that is required
to satisfy the requirement of the application.
In the example, the \emph{application service deployment} instantiates three virtual machines ($VM_A$, $VM_B$,
and $VM_C$) to service the deployment requests, a virtual monitor (\emph{vmonitor})  to
collect metrics about the performance of the slice and three controllers ($C_A$, $C_B$,
and $C_C$) to reconfigure the slice as needed to make the slice resilient to internal
and external events.  Note the slice is instantiated upon request and mapped to the
resources available from the underlying substrate. For example, in the snapshot shown,
$C_A$ can be an instance of $C_1$ provided by $NS_1$, whereas $C_C$ can be an instance
of $C_3$. Likewise, ${VM}_A$ and $VM_B$ can be instances created in $CB_1$, and so on. Since the slice
can be dynamically reconfigured, the cloud service manager might migrate $VM_B$ to  $CB_2$ upon detecting
that $CB_1$ is becoming overloaded. Alternatively, to reduce energy and monetary costs,
the CSM might release one of the virtual machines upon detecting that it is under-utilised.
Sophisticated algorithms are needed to reconfigure
the network topology. Imagine that packets from  $R_A$ to  $R_D$ travel through  $R_B$ and that the
path is exhibiting congestion problems. To avoid further complications, the CSM might decide to route packets through $R_C$ and leave $R_B$ out of the routing activities, including terminating the $R_B$ instance. The challenge of this reconfiguration is that it requires a coordinated
effort of the three controllers ($R_A$, $R_B$ and $R_C$). For instance, $R_B$ and $R_C$ need to
agree on the alterations of the routing tables of $R_C$ that they share. In addition, they need to
coordinate with $C_A$ so that $R_C$ is activated before $R_B$ is terminated. 

Typical examples of external events are
requests placed by clients against the running service application.
Though not explicitly shown in the figure, typical internal events likely to
impact the slice are failures of the virtual appliances and in particular
failures and congestion of the communication links. To minimise the risk
of link failures that might result in network partition that leave controllers unreachable,
the CSM deploys replications
of its controllers - the core elements of its slice. For instance, note
that $R_C$ can be configured either by $C_B$ or $C_C$.  The responsibility of the cloud
service is to reconfigure the slice with the assistance of the \emph{slice reconfiguration} to make it resilient.


\subsection{Resilience}
\label{controllerSoftwareArch}
The resilience of a resource slice supporting a service
relies on the use of SDN and NFV techniques to implement redundancy, decentralization
and reconfiguration. To achieve resilience, we need to provide fault-tolerance
mechanism at both the data plane (switches and links) and control plane (controllers,
controller to switches links and controller to controller links).  In this discussion
we focus on the control plane, that is, on the most sensitive part of the 
architecture as it serves as the foundation to make the data plane fault tolerant.

Notice that the three SDN controllers ($C_A$, $C_B$ and $C_C$) used by the CSM for reconfiguring the slices are geographically distributed. The reconfiguration of network functions is very likely to involve live migration of services and network functions implemented as virtual machines ($VM_A$ and $VM_B$ in the figure). Likewise, the reconfiguration of the communication infrastructure is very likely to involve alteration of the communication topology. In general, the execution of a reconfiguration script (for example, to update the routing tables of the switches $R_1$, $R_2$, etc.) will trigger the execution of underneath distributed algorithms among the SDN controllers. 
These distributed algorithms (protocols) will ensure that the state of the slice remains consistent after each reconfiguration.
The execution of these algorithms need to account for all the issues that impact asynchronous distributed systems such as the impossibility of reaching consensus in distributed systems with 
nodes that are vulnerable to crashes; and the impossibility of guaranteeing consistency, 
availability and partition tolerance (CAP theorem)~\cite{AdytyaAkella2014}. 

\begin{figure}
 \centering
 \includegraphics[keepaspectratio=true, width=\columnwidth]{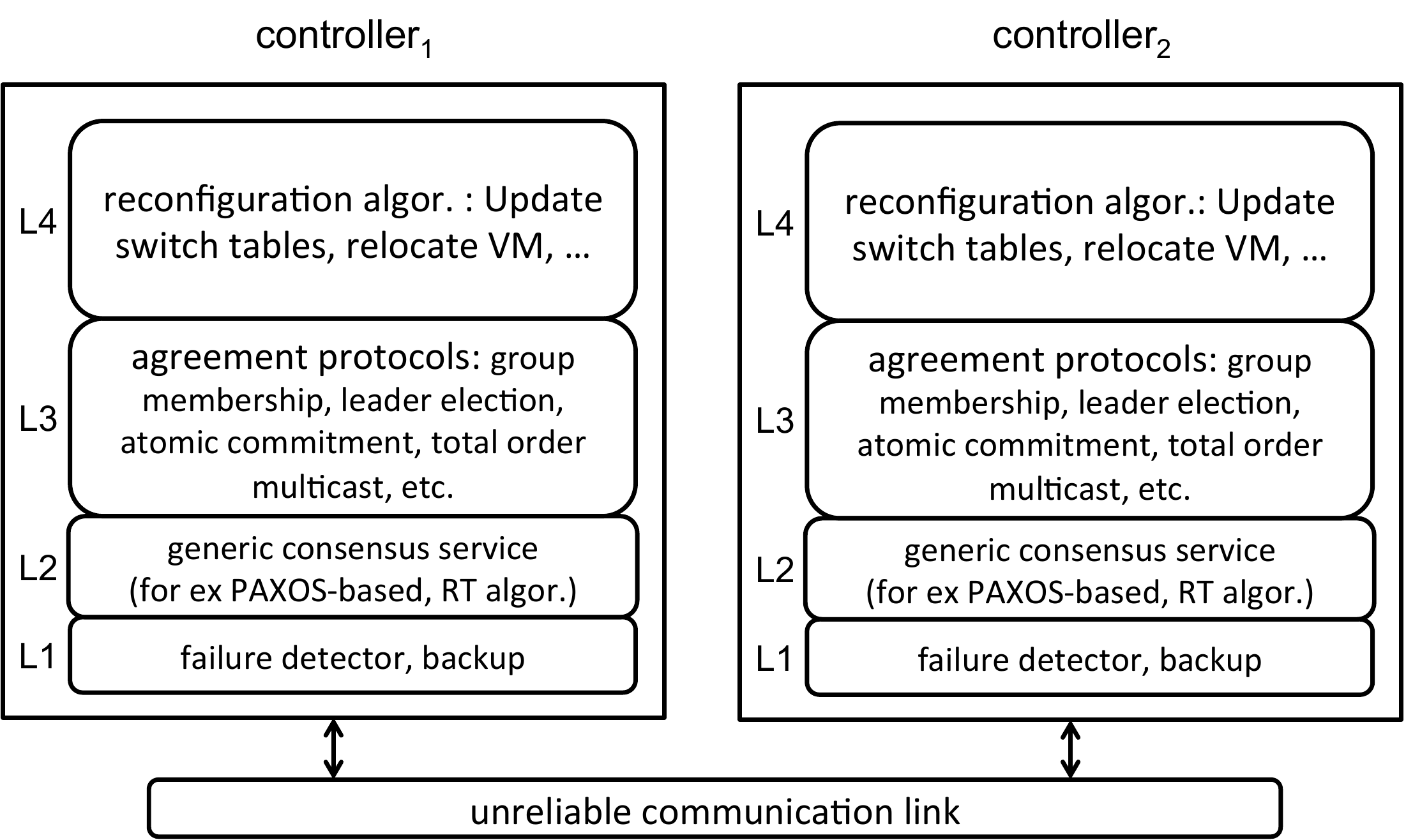}
 \caption{Controller software layers.}
 \label{fig:controllerSoftwareLayers}
\end{figure}

We anticipate different situations that will demand the execution of different reconfiguration scripts.  
To implement those scripts we follow a systematic approach based on the layered architecture shown in Figure~\ref{fig:controllerSoftwareLayers} and inspired by~\cite{Guerraoui2001}. 
A given layer offers a set of abstractions (presented as an API) to the 
layer immediately above as suggested in~\cite{Higgins:2010:INO:1859995.1860005}. 
$L$, $VM$, and $RT$ stand for Level, Virtual Machine and Real Time,
respectively. Notice that $L1$, $L2$ and $L3$ include generic distributed algorithms that 
abstracted away from the CSM but executed underneath by the 
reconfiguration scripts.


We assume that controller-to--controller communication
is performed over an unreliable network that corresponds to the unreliable communication 
links of the network substrates ($NS_1$ and $NS_2$) of Figure~\ref{fig:appServiceDeployment}.
L1 includes algorithms for failure detection and
permanent storage backup. L2 includes algorithms for achieving consensus with
different degrees of accuracy such as Paxos, real time consensus, uniform
consensus, probabilistic consensus and approximate consensus~\cite{Mostefaoui2004}.
L4 which correspond to the reconfiguration executable scripts mentioned mentioned earlier.

The modularity followed in Figure~\ref{fig:controllerSoftwareLayers}
is intended to encourage re-usability (of L1, L2 and L3 in particular) and to 
facilitate plugging in of different algorithms to match the requirements of the problem at hand.
Let us take consensus as an example. As explained in~\cite{Mostefaoui2004}, consensus algorithms with different
degrees of accuracy are the basis for implementing other distributed algorithms. 
Strong consistency may be overkill for some applications, whereas in other applications eventual consistency may 
not be enough. Regarding Figure~\ref{fig:appServiceDeployment}, a strong consensus algorithm will be executed between the controllers before altering the routing tables of $R_A$, $R_B$, $R_C$ and $R_D$ that 
define the network topology and to elect a new leader among the controllers, however, an approximate
consensus result would be enough to decide the outcome of a set of replicated monitors.
Ideally, each level from $L1$ to $L3$ should provide libraries of equivalent generic
algorithms for the benefit of the designer.

\section{Ongoing work and research challenges}
Network resilience has been the subject of research interest for decades. The effort has produced several ideas for implementing fault--tolerance mechanisms, which are mainly based on replication, design diversity and distribution of the critical software and hardware components. These mechanisms were devised before virtualisation technology (VMs, containers, unikernels, etc.) became common practice. Virtualisation opens new opportunities in network resilience but, at the same time, demands a revision of the traditional mechanisms to be assured that they cope and, if possible, take advantage of the new abstractions that virtualisation supports such as live-migration of components and seamlessly replication by means of lightweight VMs. These are open research questions. For example, it is not clear to us how conventional failure detectors will cope with dynamic instantiation, relocation and life-migration of components.
  
The use of SDN technology in the provision of network resilience raises several research questions about the implementation of the SDN control plane. The notion of the ``logically centralised controller'' that is used to operate the switches is a useful abstraction but hard to implement in practice, in particular in large networks with hundreds of thousands of switches and tens of thousands of controllers~\cite{DongtingYu2013}. The challenge is in the implementation of a distributed and scalable SDN control plane. This challenge includes the development of consensus algorithms with different degrees of accuracy that can run over different controller topologies (peer-to--peer, hierarchical, etc.) built on unreliable communication links. We are currently working in this direction. 

A close examination of the architecture of Figure~\ref{fig:appServiceDeployment} will reveal that we are dealing with a typical reactive system where a control entity is used to control the behaviour of an external entity that sends notifications to the controller about events of interest~\cite{Harel1990}. In our scenario, the control entity, the external entity and the notifications correspond respectively to the three controllers, the slice and the vmonitor. Reactive systems are notoriously hard to describe formally, reason about for correctness and implement, in particular, when they are required to respond in real time and even harder, when the response is the result of a coordinated effort of the components. The challenge here is to define notations to describe SDN topologies, that can help reason about the SDN control plane, for instance, that can help visualise the global state kept in the logically centralised controller and the partial states kept in the individual controllers.

Fault-tolerance mechanisms inevitably result in additional complexity. To avoid an explosion of complexity, we are studying the principles of structuring for fault tolerance, developed by the dependability of community~\cite{Rogeiro2009}. Intuitively, the idea is to use fault tolerance techniques systematically to ensure error containment which aims at limiting the impact of faults. In our specific scenario, error containment will ensure that failures of links or controllers do not propagate wildly and collapse the whole network. We will take a rigorous approach to quantify and compare the effectiveness of our reconfiguration actions. To give an example, to assess the resilience of a topology at the link level, we will use the concept of robustness coefficient. We will take advantage of some preliminary work that we have conducted
on the Guifi network~\cite{VegaCN}. Structuring for fault tolerance is based on the explicit separation of concerns between normal and abnormal behaviour. It requires the incorporation of mechanisms for enforcing fault models (for example, silent or byzantine crashes) and detecting errors (for example, by timeouts), as well as propagating and handling them in a structured way. These software techniques to confine errors can be supported by the architecture in the deployment of controllers. As suggested in~\cite{DongtingYu2013}, a hierarchical deployment (as opposed to peer to peer) might help to confine errors propagated by compromised or buggy controllers.

\section{Conclusions}
CNs provide an excellent scenario to deploy and use community services in a contributory or crowdsourced manner. Local CNs provide Internet access in neglected regions, but that potential is narrowed by their inherent unreliability. As they are, local CN may not reliable enough to support cloud services for business or community critical applications. We argue that reliability can be provided by means of augmenting the local community cloud  with fault--tolerance mechanisms based on the combination of monitoring and virtualisation technologies. However, the solution is not straight forward as it raises several research questions and requires the development of several supporting tool for the Cloud Service Manager. We raise some fundamental research questions in the hope of triggering discussions and motivating research in this direction.

\balance

\bibliographystyle{IEEEtran}
\small \bibliography{biblio}

\begin{thebibliography}{10}
\providecommand{\url}[1]{#1}
\csname url@samestyle\endcsname
\providecommand{\newblock}{\relax}
\providecommand{\bibinfo}[2]{#2}
\providecommand{\BIBentrySTDinterwordspacing}{\spaceskip=0pt\relax}
\providecommand{\BIBentryALTinterwordstretchfactor}{4}
\providecommand{\BIBentryALTinterwordspacing}{\spaceskip=\fontdimen2\font plus
\BIBentryALTinterwordstretchfactor\fontdimen3\font minus
  \fontdimen4\font\relax}
\providecommand{\BIBforeignlanguage}[2]{{%
\expandafter\ifx\csname l@#1\endcsname\relax
\typeout{** WARNING: IEEEtran.bst: No hyphenation pattern has been}%
\typeout{** loaded for the language `#1'. Using the pattern for}%
\typeout{** the default language instead.}%
\else
\language=\csname l@#1\endcsname
\fi
#2}}
\providecommand{\BIBdecl}{\relax}
\BIBdecl

\bibitem{Baig2015150}
R.~Baig, R.~Roca, F.~Freitag, and L.~Navarro, ``guifi.net, a crowdsourced
  network infrastructure held in common,'' \emph{Computer Networks}, vol.~90,
  pp. 150 -- 165, 2015, crowdsourcing.

\bibitem{VegaCN}
\BIBentryALTinterwordspacing
D.~Vega, R.~Baig, L.~Cerda-Alabern, E.~Medina, R.~Meseguer, and L.~Navarro, ``A
  technological overview of the guifi.net community network,'' \emph{Computer
  Networks}, vol. 93, Part 2, pp. 260 -- 278, 2015. [Online]. Available:
  \url{//www.sciencedirect.com/science/article/pii/S1389128615003436}
\BIBentrySTDinterwordspacing

\bibitem{Vega2012}
D.~Vega, L.~Cerda-Alabern, L.~Navarro, and R.~Meseguer, ``{Topology patterns of
  a community network: Guifi.net},'' in \emph{1st International Workshop on
  Community Networks and Bottom-up-Broadband (CNBuB 2012), within IEEE WiMob},
  Barcelona, Spain, Oct. 2012, pp. 612--619.

\bibitem{PAM17}
E.~Dimogerontakis, R.~Meseguer, and L.~Navarro, \emph{Internet Access for All:
  Assessing a Crowdsourced Web Proxy Service in a Community Network}.\hskip 1em
  plus 0.5em minus 0.4em\relax Cham: Springer International Publishing, 2017,
  pp. 72--84.

\bibitem{Selimi2015}
M.~Selimi and et~al., ``Cloud services in the guifi.net community network,''
  \emph{Computer Networks}, vol. 93, Part 2, pp. 373 -- 388, 2015, community
  Networks.

\bibitem{Selimi2017}
M.~Selimi, L.~Cerda-Alabern, M.~Sanchez-Artigas, F.~Freitag, and L.~Veiga,
  ``Practical service placement approach for microservices architecture,'' in
  \emph{2017 17th IEEE/ACM International Symposium on Cluster, Cloud and Grid
  Computing (CCGRID)}, May 2017, pp. 401--410.

\bibitem{Neumann2015308}
\BIBentryALTinterwordspacing
A.~Neumann, E.~Lopez, and L.~Navarro, ``Evaluation of mesh routing protocols
  for wireless community networks,'' \emph{Computer Networks}, vol. 93, Part 2,
  pp. 308 -- 323, 2015, community Networks. [Online]. Available:
  \url{http://www.sciencedirect.com/science/article/pii/S1389128615002522}
\BIBentrySTDinterwordspacing

\bibitem{Cloudy15}
R.~Baig, R.~Carbajales, P.~Escrich, J.~Florit, F.~Freitag, A.~Moll, L.~Navarro,
  E.~Pietrosemoli, R.~Pueyo, M.~Selimi, V.~Vlassov, and M.~Zennaro, ``The
  cloudy distribution in community network clouds in guifi.net,'' in
  \emph{Integrated Network Management (IM), 2015 IFIP/IEEE International
  Symposium on}, May 2015.

\bibitem{madhavapeddy2013unikernels}
A.~Madhavapeddy, R.~Mortier, C.~Rotsos, D.~Scott, B.~Singh, T.~Gazagnaire,
  S.~Smith, S.~Hand, and J.~Crowcroft, ``Unikernels: Library operating systems
  for the cloud,'' in \emph{ACM SIGPLAN Notices}, vol.~48, no.~4.\hskip 1em
  plus 0.5em minus 0.4em\relax ACM, 2013, pp. 461--472.

\bibitem{picasso17}
A.~Lertsinsrubtavee, A.~Ali, C.~Molina-Jimenez, A.~Sathiaseelan, and
  J.~Crowcroft, ``Picasso: A lightweight edge computing platform,'' in
  \emph{IEEE 6th International Conference on Cloud Networking (CloudNet'17)},
  2017.

\bibitem{Kreutz15}
D.~Kreutz, F.~M.~V. Ramos, P.~E. Verissimo, C.~E. Rothenberg, S.~Azodolmolky,
  and S.~Uhlig, ``Software-defined networking: A comprehensive survey,''
  \emph{Proceedings of the IEEE}, vol. 103, no.~1, pp. 14--76, Jan 2015.

\bibitem{Dimos13}
E.~Dimogerontakis, I.~Vilata, and L.~Navarro, ``Software defined networking for
  community network testbeds,'' in \emph{Proc. 9th IEEE Int'l Conf. on Wireless
  and Mobile Computing, Networking and Communications (WiMob'13)}, Oct 2013,
  pp. 111--118.

\bibitem{Algirdas2004}
A.~Avizienis and et~al., ``Basic concepts and taxonomy of dependable and secure
  computing,'' \emph{IEEE Transactions on Dependable and Secure Computing},
  vol.~1, no.~1, pp. 11--33, Jan-Mar 2004.

\bibitem{AdytyaAkella2014}
A.~Akella and A.~Krishnamurthy, ``A highly available software defined fabric,''
  in \emph{Proc. 13th ACM Workshop on Hot Topics in Networks (HotNets-XIII)},
  2014.

\bibitem{Guerraoui2001}
R.~Guerraoui and A.~Schiper, ``The generic consensus service,'' \emph{IEEE
  Transactions on Software Engineering}, pp. 29--41, 2001.

\bibitem{Higgins:2010:INO:1859995.1860005}
B.~D. Higgins and et~al., ``Intentional networking: Opportunistic exploitation
  of mobile network diversity,'' in \emph{Proc. Sixteenth Annual International
  Conference on Mobile Computing and Networking (MobiCom'10)}, 2010, pp.
  73--84.

\bibitem{Mostefaoui2004}
A.~M. et~al., ``Synchronous condition-based consensus,'' \emph{Distributed
  Computing}, vol.~17, no.~1, pp. 1--20, Feb. 2004.

\bibitem{DongtingYu2013}
D.~Yu and et~al., ``Authentication for resilience: the case of sdn,'' in
  \emph{Proc. Security and Privacy Workshops (SPW'13)}, 2013.

\bibitem{Harel1990}
D.~H. et~al., ``Statemate: A working environment for the development of working
  environment for the complex reactive systems,'' \emph{IEEE Transactions on
  Software Engineering}, vol.~16, no.~4, Apr. 1990.

\bibitem{Rogeiro2009}
R.~de~Lemos, ``On architecting software fault tolerance using abstractions,''
  \emph{Electronic Notes in Theoretical Computer Science}, vol. 236, no.~2, pp.
  21--–32, Apr. 2009.

\end{thebibliography}

%


\end{document}